# A compromised constitution of intersubjectivity from the perspective of a Weylian schema

LUO Dong

Institute of Advanced Study in Science, Technology and Philosophy

South China University of Technology

email: luodong@scut.edu.cn

**Abstract** Weyl proposed a schema for symbolic construction of scientific representation starting from subjective immediate experience, and moving toward the objective ordered manifolds of possibilities due to "rational motives" of comprehending the world in "the 'truly objective', exact, non-qualitative" way. Weyl's purely infinitesimal geometry may be viewed as an application of the schema. Weyl himself as well as various researchers have discussed the introduction of a local coordinate system as an effort to bridge the chasm between the intuitive and the mathematical continuum for this schema. I argue that the introduction of a local coordinate system only explained the "exact, non-qualitative" aspect of the "rational motives", and there is also a problem of how the intersubjective constitution is possible. The latter was not explicitly explained by Weyl, and corresponds to the "truly objective" aspect of "rational motives". I argue that, a comprised intersubjective constitution could be achieved by a strategy of introducing an additional structure to starting points for theoretical constructions. The strategy might also provide an explanation for gauge invariance in Weyl's unified field theory.

**Key Words:** Weyl, intersubjectvity, invariance, spacetime

## 1. Introduction

There have been various discussions on Weyl's pure infinitesimal geometry and unified field theory from the perspectives of the history of mathematics and of physics[1]. Weyl was deeply influenced by transcendental philosophy[2]. His philosophical discussions on a coordinate system as "the unavoidable residue of the ego's annihilation"[3] in efforts to bridge the deep chasm between the inexact intuitive continuum and the exact mathematical continuum have already been noticed by researchers[4]. However, there is also a problem of intersubjective constitution due to the gap between subjectivity and objectivity or intersubjectivity in Weyl's symbolic construction schema.

Weyl's philosophical thought starts with the realization that we do not directly grasp the physical world. Instead, we only have "an image, a vision, a phenomenon of our consciousness" about the physical world. From "the purely epistemological point of view", the starting point of constituting objective knowledge of the mathematical-physical world should be limited to one's immediate experience (the immediately given, or the intuitively given) whose content is exhibited in intuition.

The intuitively given is vague, inexact and subjective. However, there are "rational motives" that impel us to comprehend the world in "the 'truly objective', exact, non-qualitative" way[5]. According to Weyl, the introduction of a coordinate system is to bridge the chasm between the intuitively given inexactness and constructed exact symbolical representations of the world.

However, the intuitively given is first given to and only immediately given to one concrete subject but not to others. This means that there is a problem of how the intersubjective constitution is possible with a subjective starting point. Weyl is aware of this chasm, as he says that "…if I furthermore accept as data the contents of the consciousness of others on equal terms with my own, thus *opening myself to the*

---

[1] O'Raufeartaugh[2000], Goenner[2004], etc.
[2] Terms from Husserl's phenomenology can be easily found in Weyl's literature. For discussions of the relationship between Weyl's pure infinitesimal approach and transcendental philosophy, see Feist[2004], Ryckman[2005](especially chapter 5 and chapter 6), etc. However, it should be pointed out that Weyl also drew on philosophical sources other Husserl's phenomenology, as he himself acknowledged (Weyl[1955]) . Bell[2004] is a brief discussion on Weyl's divergence from Husserl. For Weyl, "…a number of Husserl's theses become *demonstrably false* when translated into the context of this analogy—something which, it appears to me, gives serious cause for suspecting them" (see, Weyl[1955]).
[3] Weyl[1918], p.94.
[4] For discussions about this chasm see Tieszen[2000], Ryckman[2005], etc.
[5] Weyl[1918], p.93.

*mystery of the intersubjective communication…*"[6]. However, the way he states this chasm is not as clear as his discussions on the chasm between the intuitive and the mathematical continuum.

I will argue that intersubjective constitution in Weyl's schema is partly achieved by the introduction of additional structures to starting points for theoretical constructions.

## 2. Weyl's pure infinitesimal approach

### 2.1 Weyl's schema

To build a *secure* foundation for the mathematical-physical mode of scientific cognition rather than "more or less arbitrarily axiomatized systems"[7], Weyl takes *the intuitively given* as "*the starting point*" and "the ultimate foundation" of scientific thought[8]. In scientific cognition, the intuitively given is an Archimedean point whose meaning is exhibited in intuition. "All knowledge, while it starts with intuitive description, tends toward symbolic construction"[9].

On the one hand, the contents of the intuitively given are "not composed of the mere stuff of perception, as many Positivists assert"[10], but are what an object is actually presented to a subject. They are immediate experiences of reality. They are inexhaustible and the residuum of the acts of phenomenological reductions.

On the other hand, the symbolic representation is constituted based on the intuitively given by intentional acts of the constitution. Transcendent objects in a symbolic representation only have a phenomenal existence, and can be reduced to the intuitively given by the acts of reflections (phenomenology reductions). Transcendent objects are constituted from immediate experiences of reality to represent something.

---

[6] Weyl[1927], p.117.
[7] Weyl[1918], p.24.
[8] In phenomenological terms, the immediately given serves both as the starting point of constitution or symbolic constructions of the transcendent and as the ending point of phenomenological reduction.
[9] Weyl[1927], p.75.
[10] Weyl[1918], p.4.

Weyl, following Brentano, calls them "intentional objects"[11]. The schema for symbolic construction is the following.

On the most fundamental level, the starting point consists of the following:

- "*The basic categories*" of individual objects ($x$, $y$, etc.).

- Immediately exhibited "*properties*" of and "*the primitive relations*" between objects of the basic categories ($P(x)$, $R(xy)$, etc.).

- The "*identity relation*" ("$x$ is identical to $y$") of objects in the basic categories.

Judgments that correspond to individuals' immediately given properties and relations are called "*simple or primitive judgments*". They are atomic sentences or statements in Weyl's constructions.

With the intuitively given starting point, "*complex judgments*" of "*the derived relations*" or "*the derived properties*" can be constructed by repetitively applying (in any order) "*the principles of combinations of judgments*" which define the logical functions (negation, identification of variables, conjunction, disjunction, filling in, existential quantification) of the notions of "not", "identified", "and", "or" and "there is", and the operation of "filling in" on primitive judgments in an exact way.[12]

There are ambiguities about the starting point[13], though the schema above seems clear. What are the basic categories, and primitive properties of and primitive relations between objects?[14] Weyl assumes that, in mathematical-physical cognition, "at least one category of objects underlines our investigation and that at least one of these underlying categories is that of the natural numbers". A natural number as an individual of the basic category of natural numbers and the "immediate successor" relation $S(xy)$ between two natural numbers $x$, $y$ as the primitive relation are exhibited in intuition. "A priori intuition of iteration [repetitive operations] and of the sequence of natural numbers" gives rise to our grasp of basic concepts of set theory and pure number theory without anything further being added. If there are more basic categories, a judgment about their primitive or derived relations is affiliated with their respective categories of objects. [15]

---

[11] See the Introduction of Weyl[1921].
[12] For details, see chapter 1 of Weyl[1918]. Chapter 9 of Mancosu [2010] is a discussion on this from the perspective of the foundations of mathematics.
[13] There are other ambiguities in Weyl's thought, especially in the philosophical. A similar comment can be found in Feist's[2004] efforts to clarify Weyl's concept of intuition.
[14] Furthermore, why the basic categories and primitive relations cannot be *conventionally chosen*, but are immediately *given*?

**2.2 The introduction of a local coordinate system**

**2.2.1 The chasm between the intuitive and mathematical continuum**

In Weyl's schema, the intuitively given relates to individuals[16]. "A basic category" is a category of individual objects. A natural number is an individual in the category of natural numbers. A point is an individual in a mathematical continuum. A spacetime coincidence is an individual in the spacetime continuum.

For Weyl, the basic category is supposed to be intuitively given. In the category of natural numbers, a natural number is exhibited in intuition. However, though a point (and "spatio-temporally coinciding") seems to have a very clear meaning and could serve as an individual in the basic category of the mathematical continuum, we actually never have direct experiences of a point at all. What's immediately experienced by us is the intuitive continuum, such as the intuitive space or time which consists of temporal spans or spatial extensions which are flows of our experience of time or space, or flows of consciousness. "For example, I *see* this pencil lying before me on the table throughout a certain period of time". The experience of the intuitive continuum is "reasonable and well-grounded". "Exact time- or space-points are not the ultimate, underlying atomic elements" of our experience of time or space, but are grasped based on or constituted by the immediate experience of temporal durations

---

[15] For Weyl's discussions on natural numbers, see section 5 of Weyl [1918]. Intuition of natural numbers are related to other fundamental philosophical problems about intuition. In later discussions, Weyl said that "only when I had achieved certain general philosophical insights....I became firmly convinced (in agreement with Poincaré, whose philosophical position I share in so few other respects) that *the idea of iteration, i.e., of the sequence of the natural numbers, is an ultimate foundation of mathematical thought*…"(Weyl[1918], p.48. Emphasis in the original.)

[16] Weyl's conception of intuition is a complex of Cartesian sense of intuition as being immediately evident and Kantian sense of intuition as being immediate representations or experiences of particular objects. That's why, on the one hand, the basic category is the category of individuals, on the other hand, the basic category of individuals is immediately given.

and spatial extensions[17].

On the one hand, the concept of individual (time- and space-) points is not given to us in immediate experience. Our immediate and genuine experience of the (intuitive) continuum is vague and inexact. On the other hand, in our mathematical-physical cognition of the world, there are individual points in the mathematical and spacetime continuum, and individual points in a mathematical continuum are identical or "homogeneous". "Certainly, the intuitive and the mathematical continuum do not coincide; a deep chasm is fixed between them"[18].

The chasm between the intuitive and mathematical continuum is about how an exact symbolic or mathematical world is erected from the inexact intuitively given.

**2.2.2 A coordinate system as the necessary residue**

Although, the immediately given is vague and inexact, and is "hardly the adequate medium in which physics is to construct the external world", there are "rational motives" which impel us to comprehend the world in "the 'truly objective', exact, non-qualitative" way[19]. To bridge the chasm between the intuitive and mathematical, the intuitive continuum "must be replaced by a four-dimensional continuum in the abstract arithmetical sense."[20] It is accomplished by the introduction of a coordinate system.

Since individual points cannot be exhibited in intuition and are not the individuals which can be genuinely experienced by people, Weyl insists on starting theoretical constructions with the intuitively given basic categories of individuals. For example, to develop the mathematical representation of time, one starts with the category of time spans as the basic category. By using the continuity axiom, a system of measure of time spans is established[21]. "After choosing a definite system of

---

[17] See section 6 of Weyl[1918]. This will be further discussed in the next section. The emphasis in the citation is in the original.
[18] Weyl[1918], p.93.
[19] Weyl[1918], p.93.
[20] Weyl[1927], p.113.
[21] For discussions on Weyl's introduction of a coordinate system, see chapter 5 of Ryckman[2005], chapter 11 of Ryckman&Mancosu[2010], and Bell&Korte [2015]. For Weyl's discussions on the introduction of a coordinate system, see section 6 and section 7 of chapter 2 in Weyl[1918]. To deal with the intuitive continuum, for example,

coordinates…All geometric relations thus find their arithmetic-logic representation."[22]

> The coordinate system is the unavoidable residue of the eradication of the ego in that geometrical-physical world which reason sifts from the given using 'objectivity' as its standard—a final scanty token in this objective sphere that existence is only given and can only be given as the intentional contents of the process of consciousness of a pure, sense-giving ego.[23]

The introduction of a coordination system is an *intentional* act of consciousness by a "sense-giving ego" impelled by "rational motives".

It is "unavoidable" or *necessary* because "…only reason, which thoroughly penetrates what is experientially given, is able to grasp those exact ideas"[24] by the introduction of a coordinate system. Once a coordinate system is introduced, the inexact which is intuitively given to a genuine subject or a "sense-giving ego" is replaced or "eliminated" by exact arithmetic-logic representations in mathematical-physical cognition. "Analytical geometry as founded by Descartes is the device by which we eliminate intuitive space from constructive physics."[25]

There is a "*residue*" because the "*elimination*" "does not fully succeed, and the coordinate system remains as the necessary *residue* of the ego-extinction. It is good to remember here that in practice two- or three-dimensional point-sets are usually given by actually putting a body or a figure drawn with pencil on paper before our eyes, and not by a logic-arithmetical construction of *set-defining* properties."[26]

It is necessary to point out that, a coordinate system, as the product of efforts to bridge the chasm between the intuitive and mathematical is not directly applied to the "objective" global continuum, but only applied to an intuitive space of the contents of one consciousness (which is a tangent space). As the intuitively given are limited to

---

time, with numbers, the basic category is chosen to be the category of time spans.
[22] Weyl[1934].
[23] Weyl[1918], p.94.
[24] Weyl[1918], p.94.
[25] Weyl[1927], p.117.
[26] Weyl[1927], p.75. Emphasized by me. Similar discussions on a coordinate system as "the necessary residue" can be found in various researches on Weyl, for example, chapter 5 of Ryckman[2005], chapter 11 of Ryckman&Mancosu [2010], Bell&Korte [2015], and etc.

our immediate experiences of neighborhoods, the coordinate system introduced to the intuitive is a local coordinate system.

**2.3 Pure infinitesimal geometry and the problem of space**

In geometry and the problem of space, "…only the spatio-temporally coinciding and [basic geometrical or spatial-temporal relations within] the immediate spatial-temporal neighborhood have a directly clear meaning exhibited in intuition"[27]. Weyl's starting points for mathematical constructions are *local* relations in the immediate vicinity of a point of a manifold.

Unlike in Riemannian geometry, it is only possible to compare lengths at one and the same points in Weyl's pure infinitesimal geometry[28]. The components of a vector depend on the choice of a coordinate system. The lengths of vectors depend on the choice of the "calibration" or gauge. Besides arbitrary choice of coordinate system, one can also choose an arbitrary "calibration" or standard of length at each point, which means that, at any point, only the relative lengths, the ratio of lengths of any two vectors, but not the absolute length of any one vector, are determined. Thus, Weyl's pure infinitesimal geometry introduces a 1-dimensional group of gauge transformations besides coordinate transformations. The length defined as

$$ds^2 = g_{ik}(x)dx^i dx^k \tag{1}$$

allows an additional conformal transformation

$$g \to \bar{g}_{ik} = \lambda g_{ik} \tag{2}$$

besides coordinate transformations. Here, $\lambda$ is the additional arbitrary scalar factor or gauge function. Two metrics related by a conformal gauge transformation are equally valid. And in physics, $ds^2=0$ is the equation of light cone.

---

[27] Weyl[1931], English translations in chapter 6 of Ryckman[2005], and Mancosu & Ryckman[2010] in chapter 11 of Mancosu [2010].
[28] For Weyls' discussions on his pure infinitesimal geometry, see chapter II of Weyl[1921]. For other modern discussions on Weyl's pure infinitesimal geometry and Weylian theory of spacetime, see Pauli[1981]( pp. 192-201), Bell & Korté[2015], and etc. Goenner[2004] is also a nice introduction with discussions on various reactions to Weyl's theory.

In an arbitrary coordinate system, the change of the components of a vector at a point P under a parallel displacement (which is a mathematical operation) from P to its neighboring point P′ is defined by the relation

$$d\xi^i = -\xi^r \Gamma^i_{rs} dx^s ,\qquad(3)$$

where $\Gamma^i_{rs}$ is an affine connection. A point P of a manifold is affinely related to its neighborhood if there is a coordinate system (for the immediate neighborhood of P) such that, in it, the *infinitesimal* parallel displacement leaves components of any vector at P unchanged. Such a coordinate system at point P is called a geodesic (a geodesic coordinate system) at point P.

Due to the scalar factor of the metric, the change of the length of a vector at point P displaced to its neighboring point P′ is

$$d(g_{ik}\xi^i\xi^k) = -g_{ik}\xi^i\xi^k \varphi_j dx^j ,\qquad(4)$$

where $\varphi_j$ is a length connection or gauge field. A point P of a manifold is metrically connected to its neighborhood if there is such a choice of gauge (for the immediate neighborhood of P) that any length at P undergoes no change by congruent displacements to its infinitely near points. Such a gauge at point P is called a geodesic (a geodesic gauge) at P. Making use of (1), (2) and (3), this unique connection, Weylian connection, is given by

$$\Gamma^i_{rs} = \frac{1}{2}g^{ik}(g_{kr,s} + g_{ks,r} - g_{rs,k}) + \frac{1}{2}(g^i_r\varphi_s + g^i_s\varphi_r - g_{rs}g^{il}\varphi_l) .\qquad(5)$$

The geometrical connections are invariant, in addition to arbitrary coordinate transformations, with respect to gauge transformations

$$\overline{g}_{ik} = \lambda g_{ik}, \text{ and } \overline{\varphi}_j = \varphi_j - \frac{1}{\lambda}\lambda_{,j} .\qquad(6)$$

All global relations which are related to distant regions are constructed from the above relations (the affine connection and the metric or length connection) in infinitesimal regions (by repeating infinitesimal displacements). A pure infinitesimal geometry "which will be traced through three stages; from the *continuum*, which eludes closer definition, by way of *affinely connected manifolds*, to *metrical space*"[29] is thus erected.

## 3. The problem of intersubjective constitution

### 3.1 "The mystery of the intersubjective communication"

The introduction of a coordinate system is just one step to the symbolic construction. The ultimate starting point for theoretical construction is supposed to be the genuine immediate experience that is experienced by a concrete individual subject. Apparently, what is immediately given to a concrete specific individual subject is, in principle, not directly given to or immediately experienced by another specific individual subject. What is directly given to subject $S_1$ is not directly given to another subject $S_2$. Scientific cognition is not supposed to just constitute knowledge according to those immediately given to one unique subject such as I (that would lead to solipsism). The constituted scientific knowledge should be inter-subjective and also represent what is immediately given to others and the constitutions from their perspectives. Thus, if one insists on constituting knowledge of the objective world based on what is immediately given to her, there would be a problem of how the intersubjective constitution is possible.

Weyl realized this problem of intersubjective constitution by saying that "…if I furthermore accept as data the contents of the consciousness of others on equal terms with my own, thus opening myself to *the mystery of the intersubjective communication*…"[30]

In symbolic constructions, if subject $S_1$ accepts subject $S_2$'s contents of consciousness that is only intuitively given to $S_2$, then "the mystery of the intersubjective communication" will unavoidably arise. Hence, how intersubjective constitution is possible is an unavoidable problem in Weylian schema of mathematical-physical cognition.

---

[29] Weyl[1952], p.102. Emphases in the original.
[30] Weyl[1927], p.117. Emphasized by me.

This is the second chasm besides the chasm between the intuitive and mathematical on the way to comprehending the world in "the 'truly objective', exact, non-qualitative" way. It is a deep chasm between the experience by "I" and that by others. If the first chasm is about how it is possible that an exact symbolic world is erected from the inexact intuitively given, then the second chasm is about how it is possible that an "*objective*" representation is erected from the *subjective* immediately given. Apparently, it is impossible to erect "the 'truly objective', exact, non-qualitative" world without bridging both chasms, and the "rational motives" impel us to cross both of them.

## 3.2 "The perspective view", "the photogrammetry view", and "eliminate the body"

Weyl mentioned three possible approaches to intersubjective constitutions. The first two approaches are the so-called "perspective view" approach and the "photogrammetry" approach.

> …Instead of constructing the perspective view which a given body offers from a given point of observation, or conversely constructing the body from several perspective images, as it is done in photogrammetry, we might eliminate the body…[31]

A "perspective view" is constructed by a *specific* subject $S_1$ starting from what is only immediately given to her disregarding, *at the starting points of theoretical construction*, what those which are immediately given to others $S_i$ ($i \neq 1$) are like. A "perspective view" is a view from the subject $S_1$'s specific and unique perspective $P_1$ based on the subject $S_1$'s observations which are the products of the "encounters" between the world and the subject $S_1$'s given body at a given spatial-temporal point.

The photogrammetry approach is to construct a transcendent body or object from different "perspective views" by compositions of its images obtained from different perspectives as is done in photogrammetry, for example, constructing a 3-D model for a body by 2-D images taken from different perspectives.

---
[31] Weyl[1927], p.117.

Weyl did not explain why he disapproves of "the perspective view" approach and "the photogrammetry view" approach. Possible reasons could be the following.

"The perspective view" approach might lead to solipsism. In the perspective view approach, the relations among what are intuitively given to different egos or subjects $S_i$ (i=1,2,...) are unknown. People only have the perspective dependent views of the world which are based on interactions between $S_i$'s (i=1,2,...) respective bodies and their surroundings. If people only construct "the perspective view" of the world, probably it will eventually lead to solipsism.

"The photogrammetry view" approach is disapproved probably because "constructing the body from several perspective images" by compositions of images obtained from different perspectives actually presupposed a common world or a common object ("the body") for all subjects from different perspectives. This seems to presuppose a kind of ontology basis for intersubjectivity in "the photogrammetry view" approach, which conflicts with the Weylian transcendental schema.

Weyl attempts to propose a third approach to intersubjective constitutions by "eliminating the body".

> …we might eliminate the body and formulate the problem directly as follows: let A, B, C each represent a consciousness bound to a point body, and let K be a solid contained in their field of vision. The task is to describe the lawful geometrical connections between the three images which each one of the three persons A, B, C receives of K and of the locations of the other two persons. This procedure would be more unwieldy; in fact it be bound to fail on account of the limitations and gaps in any single consciousness as compared to the complete real world.[32]

---

[32] Weyl[1927], p.117.

However, in this third approach, the problem of intersubjectivity or "the mystery of the intersubjective communication" seems ignored. "The task" to "describe the lawful geometrical connections" is based on the premise that persons A, B ,C are treated as point bodies which have not internal structures. This is to say that person A, person B, and person C are the same except being labeled differently. However, from the perspective of the problem of intersubjectivity, person A, person B, and person C could be different internally in a "mysterious" way. Anyway, "eliminating the body" does not solve "the mystery of the intersubjective communication" by saying "eliminating the body" but is just dodging the issue.

Weyl did not provide an explicit way to deal with "the mystery of the intersubjective communication". However, I will argue that the Weylian schema actually implies a possible way out of the chasm between subjectivity and objectivity. To see this, we need to first discuss two pairs of duality in Weylian schema.

## 4. Dualities

### 4.1 The double nature of the intuitively given and of ego

The intuitive given as the starting point for theoretical constructions has a double nature: the *intuitively* given and the intuitively *given*. On the one hand, it is "*given*" because the immediately *experienced* individuals (e.g. spans) in the basic category are given to a concrete subject or ego by her interactions with the world or her surroundings. On the other hand, it is "*intuitive*" because the contents of the intuitively given, the basic category of the individuals, as well as the properties of and relations among the individuals in the basic category, are exhibited in intuition. The simple judgments are about the immediately given properties and relations. They are intuitively obtained truths and have immediate certainty. Therefore, the intuitively given is a complex of the sensible (or the sensed) and the cognitive (or the cognized).

The two aspects of the intuitively given are the dual nature of one thing because they are indissoluble according to Weyl[33]. "One has to acknowledge according to Kant and Fichte: I am originally endowed with the faculty of intuition as well as sensation. A thing can exist for me only in the indissoluble unity of sensation and intuition…"[34]

Both sensibility and cognition are facilities of an ego or subject. On the one hand, the physical and empirical body of a subject is the place where "the world met with an ego". The ego is an empirical body *in* the physical world and contributes to sense data. It "meets", "sees" or interacts with the physical world, and thus is empirical. On the other hand, the ego is an agent which provides the meanings for the intuitive given, "the primitive judgments","the complex judgments" , as well as the constituted or symbolic representations of the transcendent. An ego thus also contributes meanings for the construction starting points and the constructed, which *transcends* the physical. It's a transcendental ego.

> I myself, as transcendental ego, "constitute" the world, and at the same time, as soul, I am a human ego in the world. The understanding which prescribes its law to the world is my transcendental understanding, and it forms me, too, according to these laws…[35]

As an empirical ego in the physical world, "I" respects the laws of the physical world in which there are many other human or empirical egos like me. "I", as an empirical ego, is a concrete man "who was born of a mother and who will die"[36], and is just one of the many in the world.

As a transcendental ego, the meaning of what is intuitively given to "I" as an empirical ego is exhibited in intuition and is the ultimate starting point of scientific cognition of the world. What is intuitively given to the "I" $S_1$ as an empirical ego is only intuitively given to the "I" $S_1$ but not to other people $S_i$ ($i \neq 1$), and thus it is subjective.

---

[33] Folina[2008] thinks that, like Husserl, Weyl takes intuition as both the faculty of sensibility and the faculty of understanding. Anyway, for Weyl, the intuitively given have an indissoluble double aspects. See Weyl[1934].
[34] Weyl[1934].
[35] Husserl[1970], p.202
[36] Weyl[1955].

The empirical ego is "responsible" for "seeing" those which are intuitively *given* by the world. While the transcendental ego is "responsible" for the truths or meanings of the *intuitively* given or seen.

## 4.2 The sphere of being and the sphere of meaning

Symbolic representation is the product of repetitive operations guided by "the principles of combinations of judgments" with the intuitively given as the starting point. The intuitive is the building block for constitutions or repetitive operations. The intuitively given, the basic category together with primitive properties and relations, constitutes a composite "*sphere of operation*"[37] in which any relation between objects in the category is associated with it. It is composite because it is a complex of *a sphere of things* constituted by subjects $S_i$ from the basic category and *a sphere of meanings* constituted by the meaning of primitive properties and relations.

> The world exists only as that met with by an ego, as one appearing to a consciousness; the consciousness in this function does not belong to the world, but stands out against the beings as the sphere of vision, of meaning, of image, or however else one may call it …The objects, the subjects, and the way an object appears to a subject, I model by the point, the coordinate system, and the coordinates of a point with reference to a coordinate system in geometry…Here objects (points) and subjects (coordinate system=triplets of points) belong to the same sphere of reality. The appearances of an object, however, lie in another sphere, in the realm of number.[38]

After the introduction of a *local* coordinate system, a realm of numbers is erected. By arbitrary repetitions or *iteration* of the combination principles, a realm of constructed objects, of *sets* (or property), and *functional connections* (or relations)[39]

---

[37] At the fundamental level, "a sphere of operation" consists of the category of natural numbers and the immediately exhibited relation–the immediate successor $S(xy)$. For more discussions, see Weyl[1918], pp.24-28.
[38] Weyl[1955].
[39] A set corresponds to each property. A functional connection corresponds to each relation. See section 4 of Weyl[1918].

is erected.

It is necessary to point out that, though the intuitive given provides the knowledge with certainty, the products (the symbolic construction) by repetitive operations on the intuitively given are not necessarily as real or secure as the intuitively given. Any intuitively given has its duration. The symbolic operations, though starting with the intuitively given, are made independent of the intuitively given and its duration "by being shifted on to the representing symbols which are time resisting and simultaneously serve the purpose of preservation and communication."[40] The realm of constructed objects, properties, and relations erected by mental operations is *a realm of possibilities*, or, more precisely, "an ordered manifold of possibilities producible according to a fixed procedure and open into infinity"[41].

## 5. Intersubjective constitution

### 5.1 "Rational motives" for "the objective"

The introduction of a coordinate system is considered to be impelled by "the rational motives" for comprehending the world in "the 'truly objective', exact, non-qualitative" way. However, introducing a coordinate system seems to just reflect the "exact, non-qualitative" aspect of "the rational motives". "The truly objective" aspect of "the rational motives" is not automatically accomplished by the introduction of a local coordinate system. The problem of intersubjective constitution arises because the starting point for construction appears to be subjective while there are "rational motives" that demand the intersubjective.

Scientific symbolic representation is an intersubjective symbolic representation. A scientific symbolic representation should not only be an ordered manifold of possibilities in which one transcendental ego "I" $S_1$ resides, but also a manifold consisting of many other transcendental egos $\{S_i\,(i\neq 1)\}$.

---

[40] Weyl[1934].
[41] Weyl[1932].

The intersubjectivity of this ordered manifold of possibilities requires that, on it, the ego "I" $S_1$ could "enter" another subject $S_i$'s ($i \neq 1$) intuitively given, the sphere of operation or intuitive "horizon" (intuitive space, sphere of meaning, etc.), and also other people $S_i$ ($i \neq 1$) could enter the ego "I" $S_1$'s sphere of intuitive "horizon". In other words, it means that "I" $S_1$ can appreciate the meanings or intuitively given for another subject $S_i$ ($i \neq 1$), and *vice versa*.

If intersubjectivity in this sens is achieved, in a scientific symbolic representation, one can replace $S_1$ with $S_i$ ($i \neq 1$) without damaging truths of simple and complex judgments. It means that the truth of simple and complex judgments are invariant with respect to change of subjects or consciousness.

Then, in the intersubjective ordered manifold of possibilities, my intuitively given is equivalent to others' intuitively given. In geometry and the problem of space, it means that the relations in the neighborhood of each point are equivalent. The properties of and relations in different infinitesimals are homogeneous.

> …transition to homogeneous space, where the bodily ego takes on a position on equal terms with other bodies, this is accomplished by the possibility of walking toward the distant horizon of centered space, by the free mobility of our own body in space and by the intentions of our will directed toward such motion. Not before this last step do I become capable of *imagining myself as being in the position of another person*, only this space may *he thought of as the same for different subjects*, it is a *medium necessary for constructing an intersubjective world*.[42]

"Homogeneous space" is the necessary medium for constructing an intersubjective symbolic representation of the world. It is a direct reflection of "rational motives" for "the objective".

---

[42] See Weyl[1932]. The same words can also be found in Weyl[1934]. Emphasized by me.

## 5.2 Additional structure strategy

With the mystery of the intersubjective communication, how could one make that the truth of simple and complex judgments invariant with respect to change of subjects or consciousness in scientific symbolic representations?

A specific subject $S_1$ does *not* have to *actually* communicate with another subject $S_i(i\neq 1)$ to make achieve intersubjectivity. A "*compromised*" intersubjectivity could be achieved, if subject $S_1$ could take the possible differences in subject $S_i(i \neq 1)$'s intuitively givens into consideration when subject $S_1$ is doing symbolic constructions. That is to say, if subject $S_1$ takes all possible differences contributed by the differences between her circumstances and other subjects $S_i$ 's ($i \neq 1$) circumstances into consideration, and tries to construct a scientific representation that could be *consistent with* the symbolic representations constructed by all other possible subjects $S_i(i\neq 1)$, and *if all other possible subjects $S_i(i \neq 1)$ take a similar strategy*, the $S_i$ and $S_i$ ($i\neq 1$) would reach an agreements on their symbolic constructions guided under this strategy, hence intersubjectivity achieved.

I believe the following passage from Weyl agrees with the above strategy.

> …the objective state of affairs contains all that is necessary to account for the subjective appearances. There is no difference in our experiences to which there does not correspond a difference in the underlying objective situation (a difference, moreover, which is invariant under arbitrary coordinate transformations). It comprises as a matter of course the body of the ego as a physical object.[43]

The above citation from might be interpreted as the following. For Weyl, the intuitively given impelled by the rational motives for "the objective" is equipped with possible differences of others' intuitively given. Thus, in the ordered manifold of possibilities, not only properties and relations in the distant regions are symbolic possibilities constructed by the intuitively given local one, a local region is also not only *one* specific subject $S_1$'s genuinely intuitively given but a collection of possible intuitively givens.

---
[43] Weyl[1949], p.116.

Under this strategy, a view from a given point in the manifold of possibilities is not a perspective view that is constructed by a given subject $S_1$ starting from what is only immediately given to her *disregarding* what is intuitively exhibited to others $S_i$(i $\neq$ 1). The symbolic construction is not from one subject's subjective intuitively given to an ordered manifold of possibilities from her perspective, but from a group of possible intuitively givens which reflects the "compromise" of all subjects $\{S_i\}$ to an intersubjective ordered manifold of possibilities. The "compromise" or intersubjectivity is achieve with a price. For subject $S_1$, compelled by "the rational motives" for the objective, to construct intersubjective symbolic representations, she then would need to introduce *an additional structure* to her starting points to reflect not only the circumstance of herself $S_1$ but also the circumstances of other subjects $S_i$(i $\neq$ 1).

From the perspective of the above strategy, the additional conformal structure of local relations in Weyl's pure infinitesimal geometry might be interpreted as a strategy to include all uncertainties in the calibrations, including those contributed by possible conventionality. The invariance under coordinate transformations as well as gauge transformations are, thus, necessary for the constitutions of intersubjectivity in Weyl's infinitesimal geometry.

## 6. Conclusion

The Weylian schema for symbolic constructions of scientific representations have a problem of intersubjective constitution. I argue that there is a strategy for constructing a "*compromised*" intersubjectivity in this schema by introducing an additional structure to starting points for theoretical constructions which requires every subject takes all possible differences contributed by the differences between her circumstances and other subjects' circumstances into consideration in the symbolic construction procedures. The intersubjectivity achieved by this strategy may be said to be *a priori*, contrasting with the possible *posteriori* intersubjectivity obtained after communications among subjects.

Of course, we should also keep in mind that, although the pure infinitesimal as an application of the Weylian schema is a beautiful theory, it fails to agree with empirical observations. This might agree with an observation that intersubjectivity might not mean truth. That will be another story.